\documentclass[galaxies,article,accept,oneauthor,pdftex,10pt,a4paper]{mdpi}

\firstpage{1}
\articlenumber{x}
\doinum{10.3390/------}
\pubvolume{4}
\pubyear{2016}
\copyrightyear{2016}
\externaleditor{Academic Editors: Stefano Ettori and Dominique Eckert}
\history{Received: 27 September 2016; Accepted: 9 December 2016; Published: 20 December 2016}

\pdfoutput=1

\usepackage{amssymb}
\usepackage{soul,microtype}

\Title{Using the Outskirts of Galaxy Clusters to Determine Their Mass Accretion Rate}

\Author{{Cristiano De Boni} }
\AuthorNames{Cristiano De Boni}

\address[1]{%
 Institut d'Astrophysique Spatiale, CNRS (UMR8617) Universit\'e Paris-Sud 11, B\^{a}t. 121, F-91405 Orsay, France; cristiano.deboni@ias.u-psud.fr}

\corres{Correspondence: cristiano.deboni@ias.u-psud.fr; Tel.: +x-xxx-xxx-xxxx}

\abstract{We explore the possibility of using the external regions of galaxy clusters to measure their mass accretion rate (MAR). The main goal is to provide a method to observationally investigate the growth of structures on the nonlinear scales of galaxy clusters. We derive the MAR by using the mass profile beyond the splashback radius, evaluating the mass of a spherical shell and the time it takes to fall in. The infall velocity of the shell is extracted from $N$-body simulations. The average MAR returned by our prescription in the redshift range $z=[0, 2]$ is within 20--40\% of the average MAR derived from the merger trees of dark matter haloes in the reference $N$-body simulations. Our~result suggests that the external regions of galaxy clusters can be used to measure the mean MAR of a sample of clusters.}

\keyword{galaxies clusters: general; methods: $N$-body simulations; cosmology: theory}

\begin{document}



\section{Introduction}
Recently, after the discovery of a steepening in the density profiles of halos at $\sim$0.5--1 $R_{200m}$ ({$R_{\Delta m}$ is the radius of a sphere of mass $M_{\Delta m}$ and average density $\Delta$ times the mean density of the Universe, $\rho_{m} = \rho_{c} \Omega_{m}$, where  $\rho_{c} \equiv 3H^2(z)/8 \pi G$ is the critical density of the Universe (used to define $R_{\Delta c}$) and $H$ is the Hubble parameter}) in numerical simulations \cite{dk14}, the outskirts of galaxy clusters have gathered much attention. This steepening was later recognised to correspond to the splashback radius $R_{\rm sp}$ of the recently accreted material onto the halo \cite{adc14}; {i.e.,} the outermost radius reached during the first orbit around the cluster centre. The location of the splashback radius depends on the MAR of the halo, being closer to the centre of the cluster for higher MAR. It has also been found by \cite{dk14} that for $r \gtrsim R_{\rm sp}$, the profiles are remarkably self-similar when rescaled by $R_{200m}$, while the inner profiles are more self-similar when rescaled by $R_{200c}$. For this reason, and because the splashback radius is not affected by pseudo-evolution (see \cite{dmk13} for a full explanation), it has been suggested by \cite{mdk15} to use the splashback radius as a natural physical boundary for the cluster. They found $R_{\rm sp} \sim 1.4 R_{200m} \sim 2 R_{200c}$. The existence of the splashback radius in observed galaxy clusters has been confirmed by \cite{pl16,metal16} by using {SDSS} data. Shi \cite{shi16} provided an analytical study of the outer profile of dark matter halos, along~with fitting formulae for the splashback radius. He found that the position of the splashback radius mainly depends on the MAR of the halo, but also on the amount of matter in the Universe, $\Omega_{m}$. This could be the reason why at radii larger than $R_{\rm sp}$, the profiles are more self-similar when rescaled by $R_{200m}$ than when rescaled by $R_{200c}$.

Another important radius in the outer regions of galaxy clusters that has been studied \mbox{as an alternative} definition of the splashback radius and of the halo boundary is the so-called infall radius $R_{\rm inf}$; {i.e.,} the radius where the mean radial velocity is more negative. More et al. \cite{mdk15} found that $R_{\rm inf} \approx 1.4 R_{\rm sp}$.

The mass accretion rate (MAR) of galaxy clusters has been the object of many numerical and theoretical studies \cite{vdb02,zhao03b,st04a,st04b,gio07,mcb09,zhao09,fm08,fmbk10,gts12}, but an observational measure of the MAR has never been performed. The major aim of our work \cite{dsd16} is to provide a method that can be applied to a sample of observed galaxy clusters in order to estimate their MAR.
Our method is based on the use of the cluster mass profile at large radii to determine the mass of an infalling spherical shell. The measure of the mass of the shell, along with the time it takes for the shell to infall onto the cluster, gives an estimate of the MAR.

From an observational point of view, the mass profile at large radii can be obtained through the caustic technique \cite{dg97,d99,sdmb11}. The caustic technique uses the celestial coordinates and the redshift of the galaxies to estimate, through the escape velocity, the mass profile of the cluster at very large radii---up~to several times the virial radius. Moreover, the technique determines which galaxies are members of the cluster and which are not. This method relies on the spherical symmetry of the system, but---unlike~other methods estimating the mass profile of clusters---it does not rely on any kind of equilibrium assumption. This fact is very important, since we want to study and characterise the outer regions of clusters, where equilibrium conditions are surely not satisfied.

Before applying our method to real data, we have to check whether it is capable of recovering the true MAR of objects for which the latter is well known. For this reason, we rely on $N$-body simulations of dark matter haloes and compare the MAR obtained with our method with the ``true'' MAR obtained from the merger trees for the same sample of clusters.


\section{MAR Evaluation}
\vspace{-6pt}

\subsection{The Sample}

To test our model, we use the suite of CoDECS simulations \cite{b12}. CoDECS is a set of dark matter-only simulations in different dark energy cosmologies. {{The simulated box has a comoving volume} of $(1 \ {\rm{Gpc}} \ h^{-1})^{3}$ containing $(1024)^{3}$ dark matter (CDM) particles and the same amount of baryonic particles. The mass resolution is $m_{\rm DM} = 5.84 \times 10^{10} \ {\rm{M_{\odot}}} \ h^{-1}$ for CDM particles and \linebreak$m_{b} = 1.17 \times 10^{10} \ {\rm{M_{\odot}}} \ h^{-1}$ for baryonic particles. No hydrodynamics are included in the simulation. Baryonic particles are only included to account for the different forces acting on baryonic matter in the coupled quintessence models.} As a reference model for this work, we take the $\Lambda CDM$ cosmology. Nonetheless, by constraining the growth of the structure at the nonlinear level, in principle our method could be used to discriminate among different cosmologies by constraining the growth of the structure in the nonlinear regime. So, at a different stage, we plan to directly test this hypothesis with this set of simulations by also considering the other available cosmological models.

We concentrate on massive clusters at $z=0$ and on their progenitors at higher redshifts (up to $z=2$), which is the population of objects for which we have suitable data to test our model.
Namely, we consider two samples: one sample of $2000$ clusters with median mass $M_{200c}=10^{14} \ {\rm{M_{\odot}}} \ h^{-1}$ and another sample of $50$ clusters with median mass $M_{200c}=10^{15} \ {\rm{M_{\odot}}} \ h^{-1}$, along with their progenitors. In the second sample, we are limited in the number of haloes by the size of the volume sampled by the CoDECS simulations.
Through the merger trees of the dark matter halos, we follow the evolution back in time of the progenitors of the haloes at $z=0$. In this way, for each individual halo, we can evaluate its mass accretion history (MAH); {i.e.}, the evolution with redshift of the mass of the progenitors, $M(z)$. The derivative with respect to cosmic time of the MAH is the true MAR of the halo.

\begin{figure}[H]
\centering
 \includegraphics[width=0.45\textwidth]{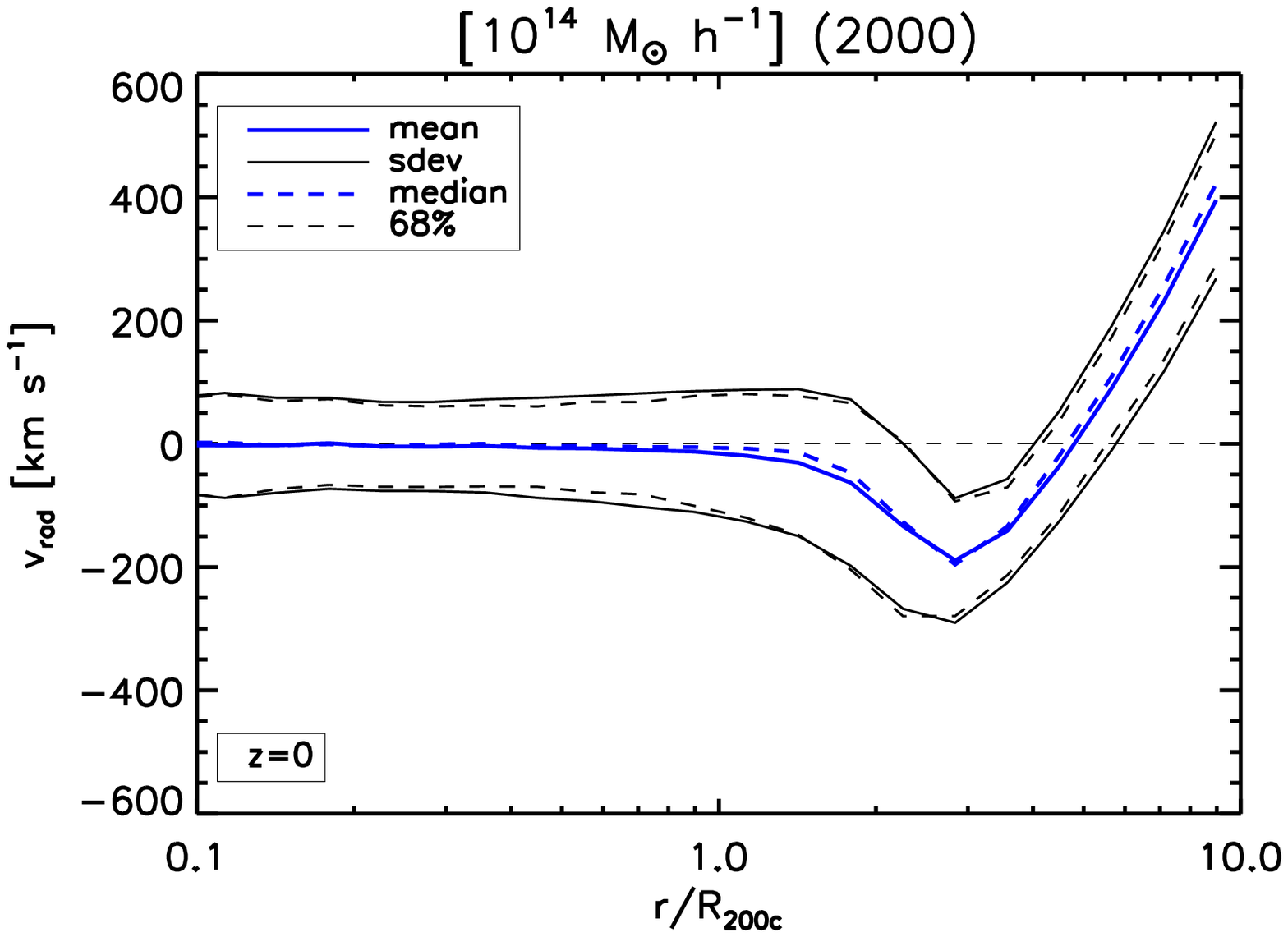}
 \includegraphics[width=0.45\textwidth]{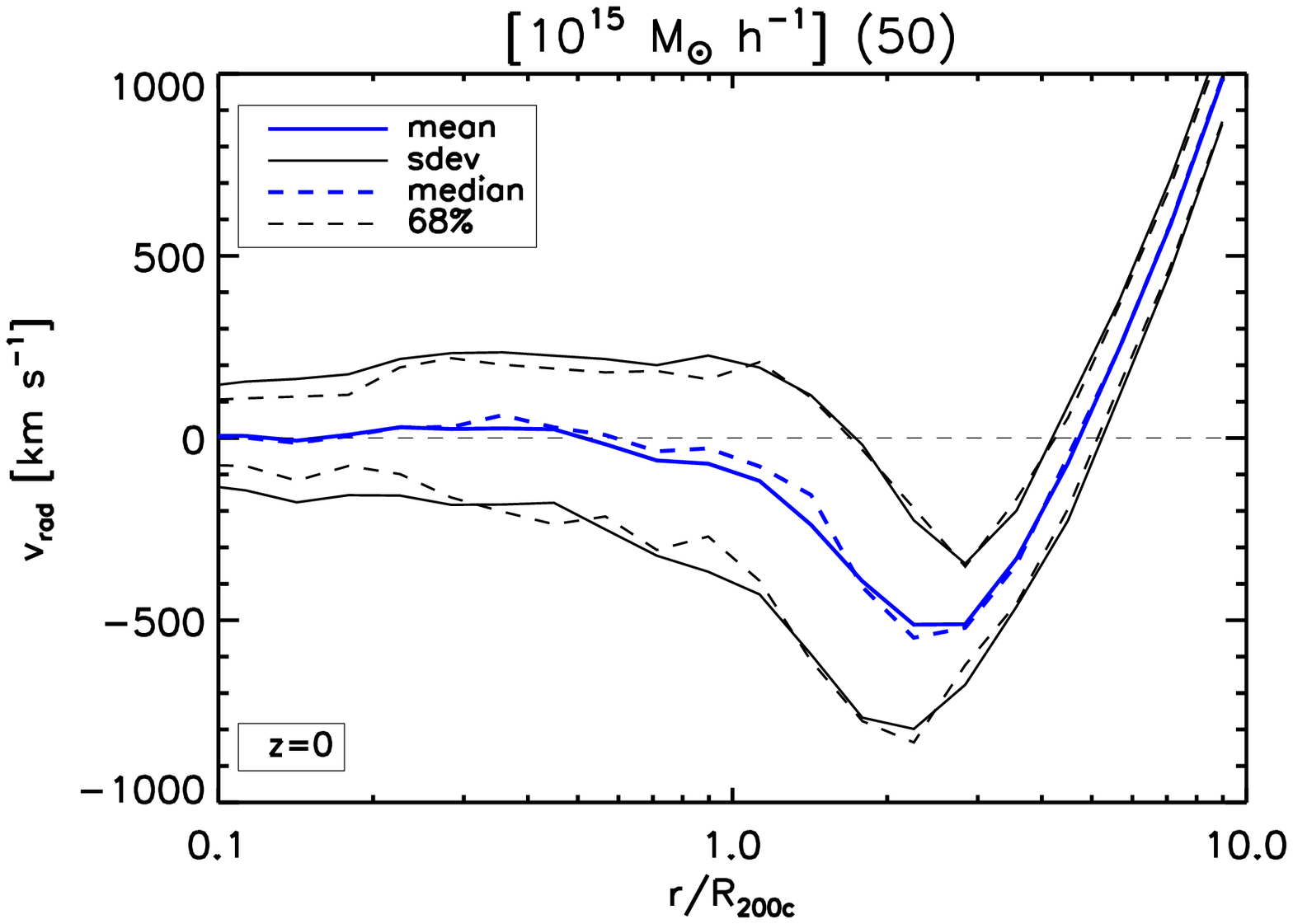}
\caption{Radial velocity profile for clusters in the $10^{14} \ {\rm{M_{\odot}}} \ h^{-1}$ (\textbf{left panel}) and in the $10^{15} \ {\rm{M_{\odot}}} \ h^{-1}$ (\textbf{right~panel}) samples at $z=0$. {{The numbers between brackets are the number of objects in the sample}}. The thick-solid and thick-dashed {blue} lines indicate the mean and median profiles, respectively. The~thin-solid and thin-dashed {black} lines indicate the standard deviation and the $68\%$ of the distribution, respectively.}
\label{radial_velocity}
\end{figure}

For each halo in the two samples at $z=0$, and for all their progenitors, we evaluate the mass profile $M(r)$ and the radial velocity profile $v_{\rm rad}(r)$ up to {$10 R_{200c}$}.
It is clear from Figure \ref{radial_velocity} that we can divide the cluster region into three zones: an inner region---($r \leq R_{200c}$) with $v_{\rm rad} \approx 0$---where~the material is orbiting around the cluster center; an infall region with $v_{\rm rad} < 0$, where the material is falling in and where $R_{\rm inf}$ is located; and an outer region with $v_{\rm rad} > 0$ dominated by the Hubble flow.

The initial infall velocity $v_{i}$ and the mass profile $M(r)$ are the only ingredients we need to provide in order to apply our model (see Section \ref{model}) to a cluster.
In this way, we evaluate the MAR with our method by using the outer mass profile and compare it with the true MAR obtained from the merger~trees.

\subsection{The Model} \label{model}

We consider a spherical cluster of radius $R_i$ and mass $M$(<$R_{i}$), surrounded by a spherical shell of matter of thickness  $\delta_{s} R_{i}$ which is falling onto it with constant acceleration $a_0 = -G M$(<$R_{i}$)/$(R_{i} + \delta_{s} R_{i}/2)^2$ {(where $G$ is the gravitational constant)} and initial velocity $v_{i}$. {By solving the equation of motion ${\rm{d}}^2 r/{\rm{d}}t^2 = a_{0}$ under these assumptions, we obtain the infall time $t_{\rm inf}$ from the equation $a_0t_{\rm inf}^2/2 + v_{i}t_{\rm inf}=-\delta_{s}R_{i}/2$, where we consider the shell to be accreted when the shell middle point (initially at  $\delta_{s} R_{i}/ 2$) reaches $R_{i}$. We can express the result as a polynomial of the shell thickness $\delta_{s}$:}
\begin{equation}
\delta_{s}^3 \frac{R_{i}^3}{4} + \delta_{s}^2 \left( R_{i}^3 + \frac{R_{i}^2}{2} v_{i} t_{\rm inf} \right) + \delta_{s} \left( R_{i}^3 + 2 R_{i}^2 v_{i} t_{\rm inf} \right) = GM(<R_{i}) t_{\rm inf}^2 - 2R_{i}^2 v_{i} t_{\rm inf} \ .
\label{shell thickness}
\end{equation}

The model has three free parameters, namely: (1) the infall velocity $v_{i}$; (2) the infall time $t_{\rm inf}$; (3)~the shell thickness $\delta_s$.
It is a convenient choice to fix the infall time to $t_{\rm inf} = 1~{\rm Gyr}$. If we do so, we~can derive the shell thickness once we know the infall velocity of the shell. We take this value for the infall time for two reasons: first of all, it is equivalent to the dynamical time ({i.e.,} the time it takes the cluster to communicate within itself through its gravitational potential) for the clusters in our analysis in the mass and redshift ranges considered; secondly, this time is also compatible with the time separation from one redshift to the other for which we have clusters in the simulations. This fact allows us to compare two MARs that are obtained from the same infall time.

In the end, $v_{i}$ is the only model parameter  that is taken from $N$-body simulations. Moreover, to minimize the fine-tuning of the model, for each sample we assign {the same initial velocity $v_{i}$} at all the clusters at a~given redshift;
~namely, the mean $v_{i}$ of all the clusters at that redshift in the sample. In this way, on one hand, we lose specific information on the individual clusters, but on the other hand, we ensure that we can apply our model to a wider class of objects---even without knowing their individual infall velocity. All of the remaining ingredients can be directly measured by~observations.

According to the recent findings about the splashback radius, we define the boundary of our cluster to be $R_{i}=2 R_{200c} \approx R_{\rm sp}$ and---for all the objects in a sample with a given redshift---we take as initial velocity the value of the mean radial velocity in the radial bin $[2-2.5]R_{200c} \approx R_{\rm inf}$; {i.e.}, $v_{i} \approx v_{\rm rad}(R_{\rm inf})$. In this way, we are sure we are considering a shell of material which has not already orbited around the cluster, but is falling in for the first time.

For each cluster at $z=0$ in our samples and for each progenitor, we evaluate the mass of the falling shell and the individual MAR as
\begin{equation}
MAR = \frac{M_{\rm shell}}{t_{\rm inf}} \ .
\label{MAR_recipe}
\end{equation}

 In the end, for each cluster, we have the evolution of the MAR with redshift. Then, for each redshift for which we have progenitors, we evaluate the mean and median MAR, so that we end with a~mean and median MAR as a function of redshift for both samples.
We compare these mean and median MARs with the true ones obtained from merger trees.

\section{Results}

\begin{figure}[H]
\centering
 \includegraphics[width=0.45\textwidth]{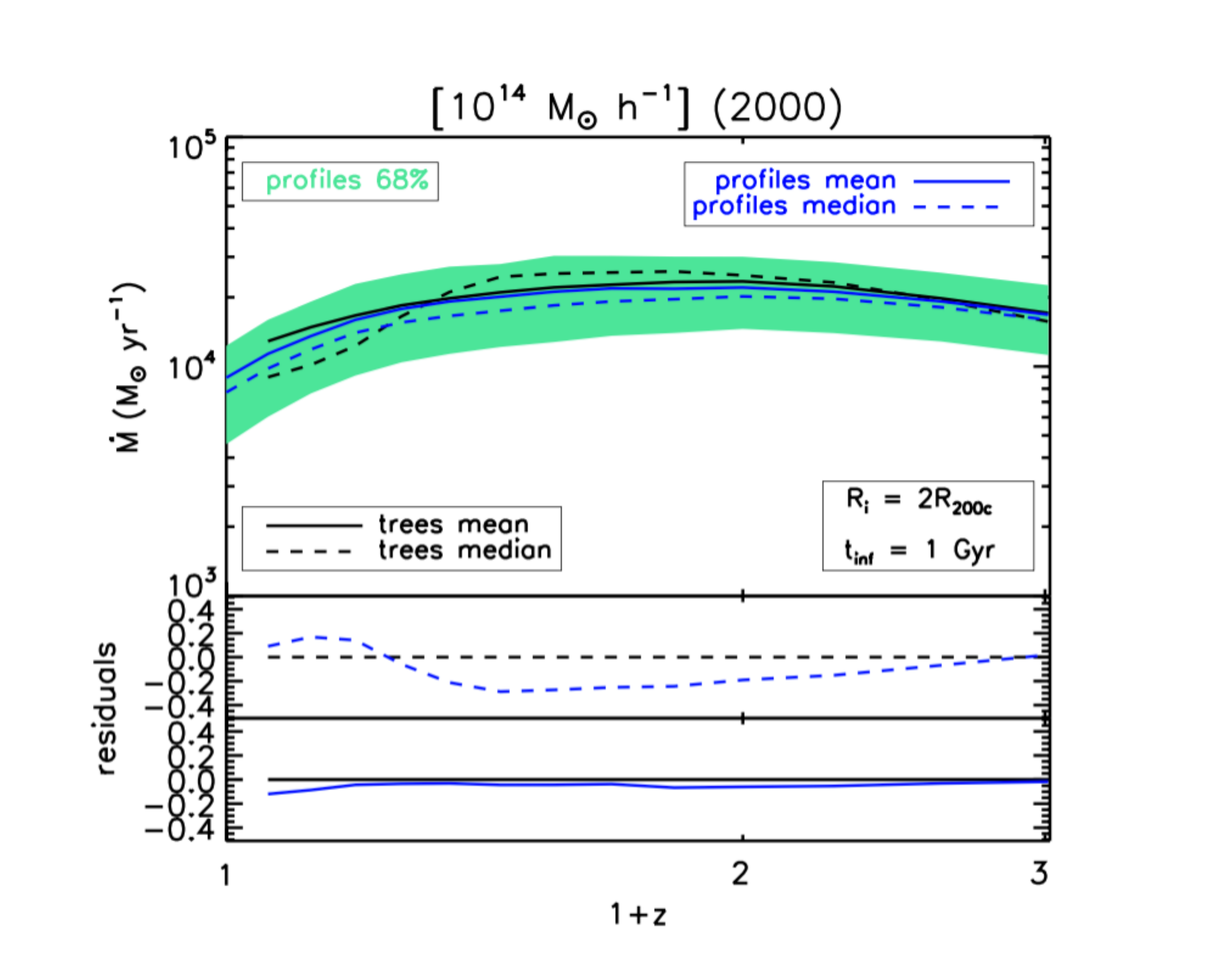}
 \includegraphics[width=0.45\textwidth]{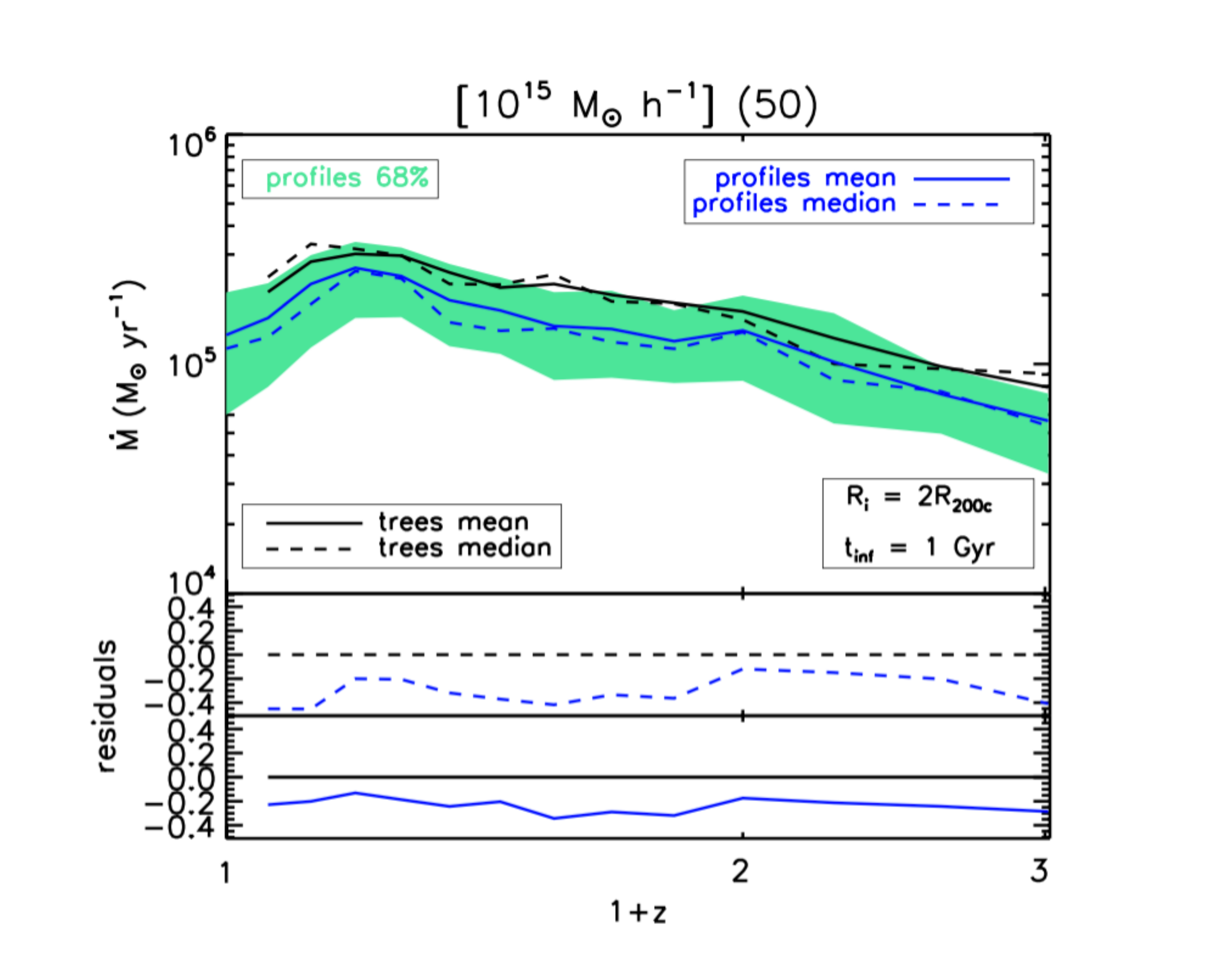}
\caption{Results of our spherical infall model for the $10^{14} \ {\rm{M_{\odot}}} \ h^{-1}$ (\textbf{left panel}) and $10^{15} \ {\rm{M_{\odot}}} \ h^{-1}$ (\textbf{right~panel}) samples and comparison with the MAR from merger trees.
The blue solid and dashed lines are the mean and median MAR from Equation (\ref{MAR_recipe}). The green area indicates the $68\%$ of this MAR distribution. The mean and median MAR from the merger trees are indicated by the black solid and dashed lines. {Residuals from the median and mean values are shown in the bottom panels}.}
\label{spherical_infall}
\end{figure}

We show the comparison between the MAR obtained with our model and the true MAR obtained from merger trees in Figure \ref{spherical_infall}.
For the $10^{14} \ {\rm{M_{\odot}}} \ h^{-1}$ sample, the true MAR from the merger trees is well within the $68 \%$ percentile of the distribution of the MARs obtained from Equation (\ref{MAR_recipe}). Moreover, our~recipe is capable of recovering the mean and median MAR from the merger trees with an accuracy of $20 \%$ in {the entire}
~redshift range considered, without any systematic trend with redshift.
For the $10^{15} \ {\rm{M_{\odot}}} \ h^{-1}$ sample, the true MAR still stays within the $68 \%$ of the distribution of our MARs in almost {the entire}~redshift range. In this case, however, the accuracy is at the level of $40 \%$, with a systematic underestimate of our mean and median values with respect to the true MAR.
We attribute this underestimate to a~statistical fluctuation due to the small number of haloes ($50$ vs. $2000$) that we have in the most massive sample. This causes the spread in the initial velocity $v_{i}$ to be higher than in the other case. Thus, in~this case, the fact of using the same $v_{i}$ for all the clusters can have a major impact in the evaluation of the MAR. Indeed, the spread of the distribution of $v_{\rm rad}$ depends on $1/\sqrt{N}$, where $N$ is the number of clusters in the sample. For the $10^{14} \ {\rm{M_{\odot}}} \ h^{-1}$ sample, the $1 \sigma$ spread of $v_{\rm rad}$ is less than $2 \%$ and propagates into a relative spread of the MAR of at most $5 \%$, while for the $10^{15} \ {\rm{M_{\odot}}} \ h^{-1}$ sample, these numbers increase up to $14 \%$ and $35 \%$, respectively.

\section{Discussion}

Our results confirm that we have developed a method to evaluate the MAR of a sample of galaxy clusters provided that we know the mass profile at radii larger than the splashback radius. The~mass profile at such large radii can be obtained from redshift surveys through the caustic technique. The~only parameter that needs to be taken from an external dataset is the initial infall velocity of the shell. We~have seen that the impact of the choice of this parameter strongly depends on the number of clusters in the sample. The larger the size of the sample, the lower the spread in the distribution of $v_{i}$ and {thus}
~the lower the effect on the obtained MAR.

The mass accretion onto individual galaxy clusters is a highly stochastic, complex, and anisotropic phenomenon. Its average behaviour has been investigated through $N$-body simulations \cite{vdb02,zhao03b,st04a,st04b,gio07,mcb09,zhao09} and semi-analytical models \cite{vdb02,fm08,fmbk10,mcb09,gts12}. Of course, our spherical infall model is not suited for recovering the MAR of individual haloes. Nonetheless, despite the simplicity of this model, it is capable of recovering the mean MAR of a sample of galaxy clusters. {The impact of  anisotropic collapse and halo triaxiality should be investigated, because it could affect the determination of the MAR of individual halos. However, since we are mainly interested in measuring the MAR of a sample of clusters, this should not be an issue in applying our method, because individual asphericities are averaged out.}

This study is only the first step towards the measurement of the MAR of a sample of observed galaxy clusters.
We plan to apply our method to the CIRS \cite{rd06} and HeCS \cite{rgdk13} catalogs, which contain clusters with masses that are comparable to the ones studied in this work.

The intermediate step before doing that is to test the impact of the caustic technique on the determination of the MAR. In other words, up to now we have considered the cluster mass profile obtained by spherically averaging the 3D particle distribution in the simulation.
We now need to apply the caustic technique to mock catalogs of the same clusters and to use the mass profile obtained in this way to derive our MAR. This will enable us to test the impact of projection effects on the determination of the MAR and to quantify all the systematics related to this method.

With the measure of the MAR of observed galaxy clusters, we can constrain the growth of structure in the nonlinear regime of galaxy clusters. In principle, this is a cosmological test that can be used to~distinguish among different cosmological models.







\vspace{6pt}


\acknowledgments{{I thank the anonymous referees for the useful comments that helped improving the presentation of this work.} I thank my collaborators Ana Laura Serra, Antonaldo Diaferio, Marco Baldi and Carlo~Giocoli for allowing me to show results from our common projects. I thank Aaron Ludlow, Margherita~Ghezzi, Giulio Falcioni, and Andrea Vittino for useful discussions. I thank Justine Brisy for additional material. I acknowledge partial support from the grant Progetti di Ateneo/CSP TO$\_$Call2$\_$2012$\_$0011 ``Marco~Polo'' of the University of Torino, the INFN grant InDark, and the grant PRIN$\_$2012 ``Fisica Astroparticellare Teorica'' of the Italian Ministry of University and Research. This work was partially supported by grants from R\'egion~Ile-de-France.}


\conflictofinterests{The author declares no conflict of interest.}

\abbreviations{The following abbreviations are used in this manuscript:\\

\noindent
\begin{tabular}{@{}ll}
MAR& Mass accretion rate\\
MAH& Mass accretion history
\end{tabular}}



\bibliographystyle{mdpi}

\renewcommand\bibname{References}



\end{document}